\documentclass[12pt]{iopart}

\begin{document}

\title{
Phenomenological approach to the critical dynamics of the QCD phase transition revisited
}

\author{Tomoi Koide\footnote{Present Address:Instituto de Fisica, Universidade de Sao Paulo, C.P.66318, 05315-970, Sao Paulo-SP, Brazil}}
\address{Institute f\"ur Theoretische Physik, J.~W.~Goethe-Universit\"at, 
D-60054 Frankfurt am Main, Germany}

\ead{koide@th.physik.uni-frankfurt.de,koide@fma.if.usp.br}

\begin{abstract}
The phenomenological dynamics of the QCD critical phenomena is revisited.
Recently, Son and Stephanov claimed that the dynamical universality class of the 
QCD phase transition belongs to model H.
In their discussion, 
they employed a time-dependent Ginzburg-Landau equation 
for the net baryon number density, which is a conserved quantity.
We derive the Langevin equation for the net baryon number density, 
i.e., the Cahn-Hilliard equation.
Furthermore, 
they discussed the mode coupling induced through the {\it irreversible} current.
Here, we show the {\it reversible} coupling can play a dominant role for describing the 
QCD critical dynamics and that the dynamical universality class 
does not necessarily belong to model H.
\end{abstract}

\submitto{\JPG}

\pacs{05.10.Gg,11.10.Wx,11.30.Rd}

\maketitle

\section{Introduction}

Dynamics near the critical point can be classified 
depending on gross variables.
For example, if the critical dynamics is described by a conserved order parameter, 
the dynamical universality class belongs to model B in the classification 
scheme of Hohenberg and Halperin\cite{ref:HH}.
If it is described by a one-component conserved order parameter and 
a three-component conserved quantity, the class belongs to model H.
To see the dynamical universality class, it is important to know the 
gross variables that contribute to the critical dynamics.

Recently, 
Son and Stephanov discussed the coupled dynamics near the QCD critical point 
phenomenologically\cite{ref:SS}.
They concluded that if we ignore the mode coupling to the energy density and 
the momentum density, the dynamical universality class of the QCD phase transition 
belongs to model B, and if we take them into account, 
it is changed to model H.
However, there are possibilities that have not been discussed by them.
1) They assumed that the time evolution of the gross variables 
(the chiral condensate and the net baryon number density) 
is proportional to thermodynamic forces 
defined by the derivatives of a Ginzburg-Landau free energy in terms of the 
gross variables.
This approach will be adequate for deriving equations for non-conserved quantities, 
for example, the chiral order parameter.
Then, we obtain the time-dependent Ginzburg-Landau (TDGL) equation.
However, for conserved quantities like the net baryon number density, 
we should adopt the Cahn-Hilliard (CH) equation instead of the TDGL equation
\cite{ref:Onuki}.
2) To describe the critical dynamics of the QCD phase transition correctly, 
we should take into account the coupling of the chiral condensate and 
other gross variables, i.e., the net baryon number density, the energy density 
and so on.
Then, it is known that there exist two channels that give rise to 
the mode coupling. 
One is the irreversible coupling where the mode coupling is induced through 
irreversible currents, and the other is the reversible coupling 
induced through reversible currents
\cite{ref:HH,ref:Onuki,ref:Kawasaki,ref:Kawasaki2,ref:Kawasaki3}.
The currents defined by thermodynamic forces are irreversible currents.
Thus, Son and Stephanov discussed only the former possibility.
However, it is known that the important mode couplings are usually 
caused by not the irreversible currents but the reversible currents, 
for instance, the liquid-gas phase transition, the superfluid transition of liquid helium 
and so on.

In this paper, 
we have two purposes.
First, we derive a phenomenological equation for the net baryon number density 
by using the CH equation.
Then, the derive equation is different from that of \cite{ref:SS}.
Second, we show how the reversible current can affect 
the relaxation behavior of the chiral condensate.

This paper is organized as follows.
In Section 2, we derive the phenomenological equations of the chiral condensate 
and the net baryon number density.
In Section 3, we discuss the irreversible coupling 
of the chiral condensate and the baryon number density.
We derive the reversible coupling in the projection operator method in Section 4.
Then, we found that the chiral condensate is coupled to the chiral current.
In Section 5, we show that the reversible coupling can change the critical dynamics.
Summary and conclusions are given in Section 6.

\section{Phenomenological equations based on the TDGL and CH equations}

We adopt the same Ginzburg-Landau 
free energy $f$ as used by Son and Stephanov\cite{ref:SS},
\begin{eqnarray}
F[\sigma,n]
= \int d^3 {\bf x}
[
\frac{a}{2}(\partial_i \sigma )^2 + b\partial_i \sigma \partial_i n 
+ \frac{c}{2}(\partial_i n)^2 + V(\sigma, n)
],
\end{eqnarray}
where $\sigma$ and $n$ denote the fluctuations of the chiral condensate and 
the net baryon number density, that is, 
$\sigma = \bar{q}q - \langle \bar{q}q \rangle_{eq}$ 
and $n = \bar{q}\gamma^0 q - \langle \bar{q}\gamma^0 q \rangle_{eq}$, 
respectively.
The potential term $V(\sigma,n)$ is given by
\begin{eqnarray}
V(\sigma,n) = \frac{A}{2}\sigma^2 
+ B\sigma n + \frac{C}{2}n^2 + {\rm higher~order~terms}.
\end{eqnarray}
The parameters $A,B,$ and $C$ satisfy the relation $AC = B^2$ at the critical point.

The thermodynamic forces are defined in the following way.
Usually, for this purpose, 
we should derive the Gibbs-Duhem relation including the chiral condensate and 
calculate the entropy production.
On the other hand, the thermodynamic forces are assumed to be defined by 
the derivatives of the free energy in terms of $\sigma$ and $n$ in \cite{ref:SS}.
In this paper, we assume that the phenomenological equations of the chiral condensate 
and the net baryon number density are given by the TDGL equation and the CH equation, respectively.
From this assumption, we define the thermodynamic forces.

First of all, we ignore the mixing between the chiral condensate and the net baryon number density.
The chiral condensate is a non-conserved quantity.
Then, the phenomenological equation of the chiral condensate is 
given by the the TDGL equation with noise,
\begin{eqnarray}
\dot{\sigma}({\bf x})
&=& -J_{\sigma}({\bf x}) + \xi_{\sigma}({\bf x}) 
= -\gamma_{ss} \frac{\delta F}{\delta \sigma} + \xi_{\sigma}({\bf x}), \label{eqn:OPE}
\end{eqnarray}
where $\gamma_{ss}$ denotes the Onsager coefficient associated with $\sigma({\bf x})$.
The last term $\xi_{\sigma}({\bf x})$ represents the noise term 
of the chiral condensate.
We do not give the concrete expression of the noise term because 
the following discussions are not affected by the noise.
On the other hand, 
the net baryon number density is a conserved quantity.
Then, the equation of the net baryon number is given by the CH equation 
with noise,
\begin{eqnarray}
\dot{n}({\bf x})
&=& -\nabla {\bf J}_n ({\bf x}) + \xi_n ({\bf x})
= -\gamma_{nn} \nabla^2 \frac{\delta F}{\delta n}+\xi_n ({\bf x}), \label{eqn:BND}
\end{eqnarray}
where $\gamma_{nn}$ denotes the Onsager coefficient associated with $n({\bf x})$ and 
$\xi_{n}({\bf x})$ represents the noise term.

Then, the currents of the chiral condensate and the net baryon number density 
are defined by 
\begin{eqnarray}
J_{\sigma} ({\bf x}) &=& \gamma_{ss} \frac{\delta F}{\delta \sigma}, \\
{\bf J}_n ({\bf x}) &=& \gamma_{nn} \nabla \frac{\delta F}{\delta n},
\end{eqnarray}
respectively.
These currents are irreversible because the time reversal symmetry is broken.
In nonequilibrium thermodynamics, 
irreversible currents are proportional to thermodynamic forces.
Thus, the two thermodynamic forces are introduced by
\begin{eqnarray}
 X_{\sigma} &=& \frac{\delta F}{\delta \sigma} = (A -a\nabla^2)\sigma + (B-b\nabla^2)n, \\
{\bf X}_{n} &=& \nabla \frac{\delta F}{\delta n} = \nabla \{ (B -b\nabla^2)\sigma + (C-c\nabla^2)n \}.
\end{eqnarray}
It should be noted that the thermodynamic force ${\bf X}_n$ defined here is vector, although 
the scalar thermodynamic force is introduced in \cite{ref:SS}.

Next, we will discuss the mixing of the chiral condensate and the net baryon number density.
The general irreversible current is given by the linear combination of the 
thermodynamic forces.
By using the two thermodynamic forces defined above, 
thus, the irreversible currents including the mixing 
of the chiral condensate and the net baryon number density are given by
\begin{eqnarray}
J_{\sigma} ({\bf x}) &=& \gamma_{ss} X_{\sigma} + \gamma_{sn} {\bf X}_{n}, \label{eqn:Jsigma} \\
{\bf J}_n ({\bf x}) &=& \gamma_{nn} {\bf X}_{n} + \gamma_{ns} X_{\sigma}. \label{eqn:Jn}
\end{eqnarray}
The irreversible current of the chiral condensate (\ref{eqn:Jsigma}) is scalar and hence the 
Onsager coefficient $\gamma_{sn}$ must be vector.
Similarly, the irreversible current (\ref{eqn:Jn}) is vector and hence the 
Onsager coefficient $\gamma_{ns}$ must be vector.
Then, by using the Onsager's reciprocal theorem, we assume 
\begin{eqnarray}
\gamma_{sn} = \gamma_{ns} = \alpha \nabla.
\end{eqnarray}
Then, the off-diagonal Onsager coefficients have the q-dependences as is discussed in \cite{ref:SS}.

However, in this paper, we adopt, what we call, the Curie principle (symmetry principle), 
where, for instance, it is forbidden that vector currents are induced by 
scalar thermodynamic forces
\cite{ref:curie,ref:prigogine,ref:prigogine2}.
For instance, 
the thermal conduction current is not induced by chemical reactions because of the Curie principle.
In our case, the irreversible current of the net baryon number density 
is not induced by $X_{\sigma}$.
Accordingly, because of the Onsager's reciprocal theorem, 
the off-diagonal Onsager coefficients $\gamma_{ns}$ and $\gamma_{sn}$ vanish.
Finally, the phenomenological equations are given by 
\begin{eqnarray}
\dot{\sigma}({\bf x})
&=& -\gamma_{ss} [ (A -a\nabla^2)\sigma ({\bf x}) + (B-b\nabla^2)n ({\bf x}) ] 
+ \xi_{\sigma}({\bf x}), \label{eqn:OPE} \\
\dot{n}({\bf x})
&=& -\gamma_{nn} \nabla^2 [ (B -b\nabla^2)\sigma ({\bf x}) + (C-c\nabla^2)n ({\bf x}) ] 
+ \xi_n ({\bf x}). \label{eqn:BND}
\end{eqnarray}
Because of the definition of the Ginzburg-Landau free energy, 
the equations still contain the cross terms that give rise to the mode coupling between the 
chiral condensate and the net baryon number density.

It should be noted that if we ignore the coupling between $\sigma$ and $n$, 
it is easy to solve the linear equation and we can see that 
the relaxation time of the chiral condensate at the vanishing momentum is given by $(\gamma_{ss}A)^{-1}$.
This result is consistent with the result of the van Hove theory\cite{ref:VH}.
If the van Hove theory is exact, 
the dynamical critical exponent is completely decided by the temperature dependence of $A$, and hence 
takes the same value as the static critical exponent.
However, the experimental data shows the deviation from the static critical exponent.
The deviation comes from the couplings between the order parameter and other gross variables, 
as we will see below.

\section{Irreversible coupling in the QCD phase transition}

In this section, we discuss the mode coupling induced by the irreversible current.

After Fourier transformation,  
the Langevin equations in the limit of ${\bf q} \rightarrow 0$ are given by 
\begin{eqnarray}
\dot{\sigma}({\bf q})
&=& ( -\gamma_{ss} A -\gamma_{ss} a q^2 ) \sigma ({\bf q}) + ( -\gamma_{ss} B - \gamma_{ss} b q^2 ) n({\bf q})
+ \xi_{\sigma}({\bf q}), \label{eqn:OPE2} \\
\dot{n}({\bf q})
&=& 
 \gamma_{nn} B q^2  \sigma({\bf q})
+  \gamma_{nn} C q^2 n({\bf q}) + \xi_n ({\bf q}). \label{eqn:BND2}
\end{eqnarray}
Here, we ignored the terms proportional to $q^4$.

Following the discussion of Son and Stephanov, we 
calculate the eigenmodes of the coupled equation.
The eigenmodes are calculated by solving the equation
\begin{eqnarray}
&& \left|
\begin{array}{cc}
-\gamma_{ss} A  -\gamma_{ss} a q^2 + i\omega~~ &
-\gamma_{ss} B -\gamma_{ss} b q^2  \\
 \gamma_{nn} B q^2 &
 \gamma_{nn} C q^2 + i\omega
\end{array}
\right|
=0. \nonumber \\
\end{eqnarray}
Up to second order in ${\bf q}$, we obtain the following 
two eigenmodes, 
\begin{eqnarray}
\omega_1 &=& 
-i \frac{  \gamma_{nn} (AC-B^2)}{ A }q^2, \\
\omega_2 &=& -i \gamma_{ss} A 
+i (\gamma_{nn} C  -\gamma_{ss} a )q^2 + i \frac{ \gamma_{nn} (AB-C^2)}{ A }q^2.
\end{eqnarray}
The corresponding eigenstates are 
\begin{eqnarray}
\left(
\begin{array}{c}
 B + O(q^2)\\
- A + O(q^2)
\end{array}
\right)~~~{\rm and}~~~
\left(
\begin{array}{c}
1 + O(q^2)\\
O(q^2)
\end{array}
\right), 
\end{eqnarray}
respectively.

Afterward, Son and Stephanov developed the argument that 
the chiral condensate and the net baryon number density are not 
independent variables 
near the critical point and their linear combination gives one gross variable as 
is the case with the liquid-gas phase transition.

However, if we take into account the possibility of the critical slowing down, 
the conclusion can be changed.
From the results of the eigenmodes and eigenstates, 
one can easily write down 
the solution of Eqs. (\ref{eqn:OPE2}) and (\ref{eqn:BND2}) 
at low ${\bf q}$,
\begin{eqnarray}
\ll \sigma({\bf q},t) \gg
&=& e^{-i\omega_2 t} \sigma({\bf q}), \label{eqn:ANA-OP}\\
\ll n({\bf q},t) \gg
&=& e^{-i\omega_1 t}
\left(n({\bf q}) - \frac{ B }{ A} \sigma({\bf q})
\right)  + \frac{B }{ A }
e^{-i\omega_2 t}\sigma({\bf q}), 
\end{eqnarray}
where $\ll~\gg$ denotes the expectation value for the noise with a suitable 
Gaussian weight function.
It should be noted that there is a constraint for the initial state:
$n({\bf 0}) = 0$, 
because $n({\bf q})$ 
is the fluctuations.
To represent dissipation, $\gamma_{ss}A$ and 
$\gamma_{nn}(AB-C^2)/A$ must be positive.
Furthermore, $\gamma_{ss}A $ becomes smaller as the temperature approaches 
the critical temperature 
because of the critical slowing down\cite{ref:KM1,ref:KM2,ref:KM3}.
Thus, it is not obvious whether we can conclude $i\omega_2$ is 
much larger than $i\omega_1$ near the critical point.
Then, 
the chiral condensate and the net baryon number density 
do not have a clear separation of time scales and cannot be reduced into 
one gross variable.

\section{Reversible couplings in the QCD phase transition}

In the previous section, we derive the irreversible currents of the chiral condensate 
and the net baryon number density in the framework of nonequilibrium thermodynamics.
However, in general, the time-evolutions of hydrodynamic variables are dominated by not only 
irreversible currents but also reversible currents.
The former yields dissipation and entropy is increased, and the latter yields 
macroscopic motion as in some cases of convection and entropy is not changed.
This structure is general and does not depend on whether the variable is conserved or not.
Thus, there can be other mode couplings caused by the reversible currents 
that have been ignored so far.
As a matter of fact, it is known that the reversible coupling 
plays a dominant role in the critical dynamics: 
the most important mode couplings are ordinarily caused by not irreversible currents 
but reversible currents
\cite{ref:HH,ref:Onuki,ref:Kawasaki,ref:Kawasaki2,ref:Kawasaki3}.
In this section, we derive the reversible currents that can affect the critical dynamics.

First, we consider the net baryon number density.
The equation of continuity of the net baryon number density is 
\begin{eqnarray}
\dot{n}({\bf x}) + \nabla  {\bf J}_{re}({\bf x}) = 0,
\end{eqnarray}
where the net baryon number current is given by 
\begin{eqnarray}
{\bf J}_{re}=\bar{\psi}\gamma^i \psi - \langle \bar{\psi}\gamma^i \psi \rangle_{eq}.
\end{eqnarray}
Thus, the net baryon number density can be coupled to the 
net baryon number current.
In short, the Langevin equation for the net baryon number density including the 
reversible current is given by
\begin{eqnarray}
\dot{n}({\bf x})
&=& -\nabla {\bf J}_{re} ({\bf x})
-\gamma_{nn} \nabla {\bf X}_n ({\bf x})
+\xi_n ({\bf x}).
\end{eqnarray}

On the other hand, the chiral condensate is not a conserved quantity, 
and hence we cannot use the equation of continuity.
However, this does not mean that there does not exist the reversible current.
The typical example is the critical dynamics of liquid helium\cite{ref:HH,ref:Onuki,ref:Kawasaki,ref:Kawasaki2,ref:Kawasaki3}.
In this case, we should solve the coupled equation of the order parameter, 
the entropy density and the superfluid velocity.
The order parameter and the entropy density are non-conserved quantities, but 
the phenomenological equations contain the reversible currents, one of which 
is proportional to the superfluid velocity.
Here, we derive the reversible currents coupled to the chiral 
condensate by using the projection operator method\cite{ref:Nakajima,ref:Zwanzig,ref:Mori,ref:US,ref:KM1,ref:KM2,ref:KM3,ref:TK}.

In the projection operator method, 
the equation of motion of an operator $A({\bf x})$ 
is given by
\begin{eqnarray}
\frac{d}{dt}A({\bf x})
&=& 
\int d^3 {\bf x}' d^3 {\bf x}''
(iLA({\bf x}),A^{\dagger}({\bf x}'))\cdot 
(A({\bf x}'),A^{\dagger}({\bf x}''))^{-1}\cdot A({\bf x}'',t) \nonumber \\
&&\hspace*{-1.5cm}+ \int^{t}_{t_{0}}ds \int d^3 {\bf x}' d^3 {\bf x}'' (f({\bf x},s),f^{\dagger}({\bf x}',0))\cdot 
(A({\bf x}'),A^{\dagger}({\bf x}''))^{-1}\cdot A({\bf x}'',t-s) \nonumber \\
&&\hspace*{-1.5cm} + f({\bf x},t), 
\end{eqnarray}
where the Liouville operator $L$ is defined by $L~O = [H,O]$ with 
the Hamiltonian $H$
\footnote{
Of course, we can apply the projection operator method to classical field theory.
Then, the Liouville operator is replaced by the Poisson bracket.
However, for simplicity, we treat all gross variables as operators in the definition of the reversible currents, 
and we replace them with the corresponding classical variables at the last of the calculation.
}.
The Hamiltonian introduced here is a microscopic Hamiltonian like the QCD Hamiltonian.
However, we do not specify it because the Hamiltonian is irrelevant to define the reversible currents, 
as we will see later.
Here, the first term on the r.h.s. is called the streaming term, 
the second is the memory term and the third is the noise term defined by 
$f({\bf x},t) = Qe^{iLQ(t-t_{0})}iLA({\bf x},t_0)$.
The Kubo's canonical correlation is defined by
\begin{eqnarray}
(F,~G) = \int^{\beta}_{0}\frac{d\lambda}{\beta}
{\rm Tr} \frac{1}{Z}[e^{-\beta H} e^{-\lambda H} F e^{\lambda H} G],
\end{eqnarray}
where $F$ and $G$ are arbitrary operators, and $Z = {\rm Tr}[e^{-\beta H}]$.

The streaming term represents the contribution from reversible currents.
Thus, in order to find the reversible currents coupled to the chiral condensate, 
we should find an operator $O$ that satisfies the following condition,
\begin{eqnarray}
(iL\sigma({\bf x}),O({\bf x}') ) = 
(\frac{d}{dt}\sigma({\bf x}),O({\bf x}')) = {\rm Tr}[\rho[\sigma({\bf x}),O({\bf x}')]] 
\neq 0.\label{eqn:stream}
\end{eqnarray}
First, we assume $O = \bar{\psi}\Gamma \psi$.
Then, $\Gamma = \gamma^i$ and $\gamma^0 \gamma^i$ can satisfy this condition, 
because $[\sigma({\bf x}),O({\bf x}')] \neq 0$ 
(This condition corresponds to the non-vanishing Poisson bracket in \cite{ref:HH}).
However, the net baryon number current that is given by $\Gamma= \gamma^i$ 
is the reversible current for the net baryon number density.
As a matter of fact, it is easy to calculate the expectation value in the free gas approximation.
Then, one can see that Eq. (\ref{eqn:stream}) vanishes for $\Gamma = \gamma^i$.
It follows that the net baryon number current is not directly coupled to the chiral condensate.
Thus, the reversible current of the chiral condensate can be 
\begin{eqnarray}
{\bf J}_{\sigma}({\bf x}) = \bar{\psi}({\bf x}) \gamma^0 \gamma^i \psi({\bf x})
- \langle \bar{\psi}({\bf x}) \gamma^0 \gamma^i \psi({\bf x}) \rangle_{eq}.
\end{eqnarray}

We are particularly interested in the coupling of the chiral condensate 
and the net baryon number current, because, as is discussed in \cite{ref:KM3}, 
the net baryon number density is a promising gross variable with large time scale.
However, if we put $\Gamma=\gamma^0$, one can show that 
$[\sigma({\bf x}),n({\bf x}')] = 0$.
Thus, the net baryon number density is not coupled to the chiral condensate directly.
Now, we consider nonlinear couplings and 
assume $O = (\bar{\psi}\gamma^0 \psi) (\bar{\psi}\Gamma' \psi)$.
In order to satisfy the condition $[\sigma({\bf x}),O({\bf x}')] \neq 0$, 
we should choose, for example, $\Gamma' = \gamma^0 \gamma^i$.

Finally, the Langevin equation for the chiral condensate including the 
reversible current is given by
\begin{eqnarray}
\dot{\sigma}({\bf x})
&=& -R\nabla \cdot {\bf J}_{\sigma} ({\bf x})
 -T\nabla \cdot (n({\bf x}){\bf J}_{\sigma} ({\bf x}))
-\gamma_{ss} {\bf X}_{\sigma}({\bf x}) + \xi_{\sigma}({\bf x}). 
\end{eqnarray}
It should be noted that 
this definition of the reversible coupling is only 
the necessary condition, because there is a possibility that 
the coefficients of reversible currents $R$ and $T$ vanish after calculating 
the expectation value in Eq. (\ref{eqn:stream}).
The coefficients depend on the choice of the microscopic Hamiltonian $H$ and hence 
microscopic calculations are needed.
However, in our phenomenological discussion, we leave them as unknown parameters.
After all, to describe the critical dynamics near the chiral phase transition 
including the reversible couplings, 
we should solve the coupled equation of 
$\sigma$, $n$, ${\bf J}_{n}$ and ${\bf J}_{\sigma}$.

\section{Analysis based on the mode coupling theory}

In this section, we show that the reversible coupling derived above 
can affect the relaxation of the chiral condensate and 
change the dynamical universality class of QCD.

First, we make the following ansatz for 
the chiral current;
\begin{eqnarray}
{\bf J}_{\sigma}({\bf x}) = \sigma({\bf x}) {\bf v}_{\sigma}({\bf x}).
\end{eqnarray}
We further assume that the reversible flow of the chiral condensate 
are incompressible,
\begin{eqnarray}
\nabla \cdot {\bf v}_{\sigma} = 0.
\end{eqnarray}  
Then, the Langevin equation for the chiral condensate
is given by 
\begin{eqnarray}
\dot{\sigma}({\bf x})
&=& - R({\bf v}_{\sigma}({\bf x}) \cdot \nabla) \sigma({\bf x})
- T 
({\bf v}_{\sigma}({\bf x})\cdot \nabla) (n({\bf x})\sigma({\bf x}) ) 
\nonumber \\
&&-\gamma_{ss}X_{\sigma}({\bf x}) + \xi_{\sigma}({\bf x}). \label{eqn:OPEWRC}
\end{eqnarray}
The second term on r.h.s. gives rise to the coupling of the chiral condensate and 
the net baryon number density. 
However, this is a higher order nonlinear term and we drop it for simplicity.

To estimate the effect of the nonlinear reversible coupling, 
we apply the technique developed in the mode coupling theory, 
where the nonlinear effect is calculated perturbatively\cite{ref:Kawasaki,ref:Kawasaki2,ref:Kawasaki3}
\footnote{The mode coupling theory in the chiral phase transition is discussed also 
in \cite{ref:OFO}}.
The nonperturbative solution is calculated by setting $R=0$,
\begin{eqnarray}
\sigma^0 ({\bf q},t) 
= e^{-i\omega_2 t} \sigma({\bf q}) + \int^{t}_{0} ds e^{-i\omega_2(t-s)}
\xi_{\sigma}({\bf q},s). 
\end{eqnarray}
We further introduce the time correlation function of the chiral condensate for $t \ge t' \ge 0$, 
\begin{eqnarray}
G_{\bf q}(t-t') 
\equiv \langle \sigma ({\bf q},t)\sigma ({\bf -q},t') \rangle /
\chi_{\bf q},
\end{eqnarray}
where $\chi_{\bf q} = \langle \sigma({\bf q}) \sigma ({\bf -q}) \rangle$, 
and $\langle~\rangle$ denotes the expectation value for an initial state 
including the weight function of the noise.
When we ignore the nonlinear term, the correlation function is 
\begin{eqnarray}
G^0_{\bf q}(t-t') 
\equiv \langle \sigma^0 ({\bf q},t)\sigma^0 ({\bf -q},t') \rangle /
\chi_{\bf q} = e^{-i\omega_2(t-t')}.
\end{eqnarray}

With the help of the nonperturbative solution, 
the solution of Eq. (\ref{eqn:OPEWRC}) is given by
\begin{eqnarray}
\sigma({\bf q},t) 
&=& \sigma^{0}({\bf q},t) 
- R\int^{t}_{0}ds \frac{1}{V}\sum_{\bf k}
e^{-i\omega_2(t-s)} 
(i{\bf k}\cdot {\bf v}_{\sigma}({\bf q-k},s)) \sigma({\bf k},s),  
\label{eqn:ANA-SOL}
\end{eqnarray}
Then, the time correlation function is developed by the following equation,
\begin{eqnarray}
\frac{d}{dt}G_{\bf q}(t)
= -i\omega_2 G^0_{\bf q}(t) 
 - \frac{R}{\chi_{\bf q} V}\sum_{\bf k}
 \langle ( i{\bf k} \cdot {\bf v}_{\sigma}({\bf q-k},t)) \sigma({\bf k},t)
\sigma({\bf -q}) \rangle .\label{eqn:EX-TCF}
\end{eqnarray}

Substituting Eq. (\ref{eqn:ANA-SOL}) into the second term on the r.h.s. of 
Eq. (\ref{eqn:EX-TCF}) and expanding in powers of the $R$ up to second order, 
we have 
\begin{eqnarray}
\frac{d}{dt}G_{\bf q}(t) 
&=& -i\omega_2 G^0_{\bf q}(t) 
- \frac{R}{\chi_{\bf q} V}\sum_{\bf k}
 \langle ( i{\bf k}\cdot {\bf v}_{\sigma}({\bf q-k},t) ) \sigma^{0}({\bf k},t)
\sigma ({\bf -q}) \rangle \nonumber \\
&&\hspace*{-2cm}+ \frac{R^2}{\chi_{\bf q} V^2}\sum_{\bf k,k'} \int^{t}_{0}ds e^{-i\omega_2(t-s)} 
\langle ( i{\bf k}\cdot {\bf v}_{\sigma}({\bf q-k},t) )
(i{\bf k'}\cdot {\bf v}_{\sigma}({\bf k-k'},s)) \sigma^0 ({\bf k'},s)
\sigma ({\bf -q}) \rangle . \nonumber \\
\end{eqnarray}
We further apply the decoupling approximation as follows;
\begin{eqnarray}
&& \langle ( i{\bf k}\cdot {\bf v}_{\sigma}({\bf q-k},t) ) \sigma^{0}({\bf k},t)
\sigma ({\bf -q}) \rangle \nonumber \\
&& \hspace*{0.5cm}
\approx \langle ( i{\bf q}\cdot {\bf v}_{\sigma}({\bf 0},t) ) \rangle 
\langle \sigma^{0}({\bf q},t) \sigma ({\bf -q}) \rangle 
\delta^{(3)}_{\bf k,q}, \\
&& \langle ( i{\bf k}\cdot {\bf v}_{\sigma}({\bf q-k},t) )
(i{\bf k'}\cdot {\bf v}_{\sigma}({\bf k-k'},s)) \sigma^0 ({\bf k'},s)
\sigma ({\bf -q}) \rangle \nonumber \\
&&\hspace*{0.5cm} \approx
\langle ( i{\bf k}\cdot {\bf v}_{\sigma}({\bf q-k},t) )
(i{\bf q}\cdot {\bf v}_{\sigma}({\bf k-q},s)) \rangle 
\langle \sigma^{0}({\bf q},t) \sigma ({\bf -q}) \rangle 
\delta^{(3)}_{\bf k',q}. 
\end{eqnarray}
Then, we obtain 
\begin{eqnarray}
\frac{d}{dt}G_{\bf q}(t) 
&=& -i\omega_2 G^0_{\bf q}(t) 
- \frac{R}{ V}
 \langle ( i{\bf q}\cdot {\bf v}_{\sigma}({\bf 0},t) ) \rangle G^0_{\bf q}(t)
\nonumber \\
&&\hspace*{-1.5cm}+ \int^{t}_{0}ds \frac{R^2}{ V^2}\sum_{\bf k}
G^0_{\bf q}(t-s) 
\langle ( i{\bf k}\cdot {\bf v}_{\sigma}({\bf q-k},t) )
(i{\bf q}\cdot {\bf v}_{\sigma}({\bf k-q},s)) \rangle  G^0_{\bf q}(s).\nonumber \\ 
\end{eqnarray}

This equation includes the reversible coupling 
up to second order.
To resum the second order correction, we 
replace one $G^0_{\bf q}(t)$ in each term 
with the full time correlation function 
$G_{\bf q}(t)$, 
\begin{eqnarray}
\frac{d}{dt}G_{\bf q}(t)
&=& -i\omega_2 G_{\bf q}(t) 
- \frac{R}{V} \langle ( i{\bf q}\cdot {\bf v}_{\sigma}({\bf 0},t) ) \rangle G _{\bf q}(t)
+ R^2\int^{t}_{0}ds \Xi_{\bf q}(t,s) G_{\bf q}(s),\nonumber \\ 
\end{eqnarray}
where we introduced the memory function $\Xi_{{\bf q}}(t,s)$,
\begin{eqnarray}
\Xi_{{\bf q}}(t,s)
&=&  \frac{1}{ V^2}\sum_{\bf k}
\langle ( i{\bf k}\cdot {\bf v}_{\sigma}({\bf q-k},t) )
(i{\bf q}\cdot {\bf v}_{\sigma}({\bf k-q},s)) \rangle G^0_{\bf q}(t-s).
\end{eqnarray}
This resummation corresponds to the ring-diagram approximation\cite{ref:Kawasaki,ref:Kawasaki2,ref:Kawasaki3}.
If we can employ the Markov approximation, that is, 
the memory function relaxes rapidly and vanishes at late time, 
the time dependence of the memory function is approximately replaced by 
the delta function,
\begin{eqnarray}
\Xi_{{\bf q}}(t,s) \approx 2\Xi_{{\bf q}}\delta(t-s),
\end{eqnarray}
with
\begin{eqnarray}
\Xi_{{\bf q}} = \lim_{t\rightarrow \infty}
\int^{t}_{0}ds \Xi_{\bf q}(t,s).
\end{eqnarray}

We further assume a simple diffusion equation for the dynamics of the 
velocity field and 
ignore the reversible coupling\cite{ref:Onuki}, 
\begin{eqnarray}
\left( \frac{\partial}{\partial t} - \nu \nabla^2 \right) 
{\bf v}_{\sigma}({\bf x},t)= 0.
\end{eqnarray}
Then, the correlation function of the velocity field is given by 
\begin{eqnarray}
\sum_{\bf k}\langle (i{\bf k}\cdot {\bf v}_{\sigma}({\bf q-k},t))
(i{\bf q}\cdot {\bf v}^*_{\sigma}({\bf q-k},s)) \rangle 
&=& - \frac{q^2}{3} \sum_{\bf k} \langle {\bf v}_{\sigma}({\bf k},t)
\cdot {\bf v}^*_{\sigma}({\bf k},s) \rangle \nonumber \\
&=& - \frac{q^2}{3}\sum_{\bf k}e^{-\nu {\bf k}^2 (t-s)} 
\langle {\bf v}_{\sigma}({\bf k})
\cdot {\bf v}^*_{\sigma}({\bf k}) \rangle. \nonumber \\
\end{eqnarray}
Here, the initial correlation of the velocity is assumed to be Gaussian.
Thus, we have
\begin{eqnarray}
\Xi_{\bf q} 
= -\frac{q^2}{3 V^2}\sum_{\bf k}
\frac{\langle {\bf v}_{\sigma}({\bf k})
\cdot {\bf v}^*_{\sigma}({\bf k}) \rangle}{-\nu {\bf k}^2 - i\omega_2}.
\end{eqnarray}
Finally, we obtain the renormalized damping rate of the chiral condensate,
\begin{eqnarray}
L^R = i\omega_2 
+ \frac{R}{ V} \langle ( i{\bf q}\cdot {\bf v}_{\sigma}({\bf 0},t) ) \rangle 
+ \frac{R^2 q^2}{3 V^2}\sum_{\bf k}
\frac{\langle {\bf v}_{\sigma}({\bf k})
\cdot {\bf v}^*_{\sigma}({\bf k}) \rangle}{-\nu {\bf k}^2 - i\omega_2}.
\end{eqnarray} 
Because of the critical slowing down\cite{ref:KM1,ref:KM2,ref:KM3}, 
$i\omega_2$ becomes smaller and vanishes 
as the temperature approaches the critical temperature
\footnote{This disappearance occurs in the chiral limit. 
When we have finite current quark masses, 
$i\omega_2$ can still be small but finite even at the critical temperature.}.
If the velocity field also becomes slower 
as the temperature is lower to the critical temperature, and hence the diffusion constant $\nu$ vanishes 
at the critical point as is the case with diffusion constants in 
spin diffusion processes and binary fluid mixtures\cite{ref:Onuki}, 
the third term on the r.h.s. becomes larger than the first term, 
where the effect of the irreversible coupling is included.
This is the reason why the reversible coupling can be more important than the 
irreversible coupling and the latter is ignored in various cases 
\footnote{Of course, we cannot say anything about 
the nonlinear irreversible coupling in this discussion.}.
Then, the third term plays an important role in the critical dynamics of the chiral phase transition.
This suggests that there is the possibility that the critical dynamics is changed 
by introducing reversible currents.

\section{Summary and concluding discussions}

We have discussed the phenomenological equations to describe the critical dynamics 
of the QCD phase transition.

First of all, we derived the Langevin equations without the reversible couplings.
We adopted the time-dependent Ginzburg-Landau equation for the time evolution of 
the chiral condensate and the Cahn-Hilliard equation for that of the 
net baryon number density.
However, because of the Curie principle, the off-diagonal Onsager coefficients vanish.

Next, we discussed the reversible coupling in the QCD critical dynamics.
The chiral condensate is coupled not only to the 
net baryon number density but also to the chiral current 
through the reversible coupling.
The reversible coupling 
plays an important role for the relaxation of the chiral condensate,
because of the critical slowing down and the decreasing diffusion constant 
at the critical points.
Then, the effect of the irreversible coupling 
is small\cite{ref:HH,ref:Onuki,ref:Kawasaki,ref:Kawasaki2,ref:Kawasaki3}.

It follows that the dynamical universality class can be changed 
from model H discussed by Son and Stephanov.
They claimed that, if we can ignore the coupling to 
the energy density and the momentum density, 
the universality class belongs to model B, and 
if we take them into account, it is changed to model H.

In this study, 
we found that their discussion can be changed as follows:
1) Son and Stephanov concluded that the linear combination of the 
chiral condensate and the net baryon number density plays 
the role of an order parameter 
as in the case of the liquid-gas phase transition.
However, if we take the possibility of the critical slowing down into account, 
the appearance of the reduction of the gross variables is not clear.
2) The chiral condensate can be coupled to reversible currents, i.e. the chiral current 
and this coupling can change the dynamical universality class.

If we take the reversible coupling of the chiral current into account, 
we must solve the coupled equation for the chiral condensate and 
the chiral current.
The chiral condensate is not conserved and the chiral current 
is assumed to obey the diffusion equation.
This situation is similar to the isotropic antiferromagnet, where 
the dynamical universality class belongs to model G.
In the isotropic antiferromagnet, the order parameter 
is the three-component non-conserved vector density.
The order parameter is coupled to a three-component conserved vector density that 
obeys a diffusion equation.
In the chiral phase transition, the order parameter has four component in general, 
i.e., $\sigma$ and $\pi^i$.
Therefore, we can conclude that the dynamical universality class is 
similar to model G, noting, however, that the dimension of the order parameter 
is not three but four.

In this calculation, we ignored the reversible coupling to 
the net baryon number density, the net baryon number density current and 
so on
\footnote{
The importance of conjugate fields as the candidates 
of gross variables near the chiral phase transition 
is discussed in \cite{ref:OFO}.
Furthermore, it is known that if we ignore the conjugate field, 
it sometimes happen that we cannot separate slow and fast time scales\cite{ref:KM2,ref:KM3}.
}.
When we take them into account, the dynamical universality class 
can be changed again.
Now, we have five gross variables that possibly 
couple to the chiral condensate;
the chiral current, the net baryon number density, the net baryon number current, 
the energy density, and the momentum density.
The several gross variables may not be independent 
as is the case with the liquid-gas phase transition.
To discuss the possibility, 
we need a microscopic calculation as is done in \cite{ref:KM2,ref:KM3,ref:TK}.

The criteria for choosing gross variables to describe the QCD critical dynamics 
are discussed in \cite{ref:SS,ref:RW} 
and the criteria do not allow reversible currents as gross variables.
However, 
there is no general criteria to choose 
the complete set of gross variables\cite{ref:Kawasaki,ref:Kawasaki2,ref:Kawasaki3}.
In this paper, we have assumed that the chiral current is a gross variable.
Of course, to see whether the reversible current is slow 
enough to be interpreted as a gross variable, we should calculate 
the time correlation function of the reversible current that 
characterizes the time scale of the reversible current\cite{ref:KM3}.
However, it will be natural to expect that the coarse-grained equation of the chiral condensate 
has the contributions of the reversible currents, 
because the transfer of the fluctuations of the chiral condensate 
should be induced by the macroscopic flows of quarks and gluons, even when 
the flows do not produce entropy.
As a matter of fact, the fluctuations vary even in thermal equilibrium.

We derived the Langevin equation to describe the QCD critical dynamics 
based on nonequilibrium thermodynamics, where 
the dissipation process is described by diffusion.
However, as is pointed out in many papers, 
normal diffusion equations do not obey causality;
their propagation speed is infinite and beyond the speed of light.
To avoid this difficulty, extended irreversible thermodynamics (EIT) has been 
proposed where diffusion processes have finite relaxation time
\cite{ref:ET1,ref:ET2,ref:ET3,ref:MURONGA,ref:Mohamed}.
Thus, to discuss the QCD critical dynamics, 
it may be better to use EIT 
As a matter of fact, 
the microscopic calculation of the chiral condensate shows 
dissipation exhibiting oscillation and 
seems to support extended thermodynamics\cite{ref:KM2,ref:KM3,ref:TK}.
Then, the critical dynamics in relativistic systems can be different from 
the non-relativistic cases
\footnote{Recently, the projection operator method is applied to 
derive a coarse-grained equation for a conserved quantity.
The equation is similar to the causal diffusion equation rather than the acausal 
diffusion equation\cite{ref:TK2}.}\cite{ref:KKR}.

\ack

The author are grateful to C.~Greiner, A.~Muronga and D.~Rischke 
for fruitful discussions and helpful comments on the manuscript, and 
acknowledges a fellowship from the Alexander von Humboldt 
Foundation and the financial support from FAPESP(04/09794-0).

\end{document}